\documentclass{article}

\usepackage{arxiv}

\usepackage[utf8]{inputenc} 
\usepackage[T1]{fontenc}    
\usepackage{hyperref}       
\usepackage{url}            
\usepackage{booktabs}       
\usepackage{amsfonts}       
\usepackage{nicefrac}       
\usepackage{microtype}      
\usepackage{lipsum}		
\usepackage{graphicx}
\usepackage{doi}

\usepackage{amsmath}
\usepackage{arxiv}
\usepackage{dsfont}
\usepackage{booktabs}
\usepackage{longtable}
\usepackage{multirow}
\usepackage{graphicx}
\usepackage[utf8]{inputenc}
\usepackage[english]{babel}
\usepackage{enumitem}
\usepackage{amsthm}
\usepackage[ruled,vlined]{algorithm2e}
\usepackage{accents}
\usepackage{graphicx}
\usepackage{blindtext}
\usepackage{csvsimple}
\usepackage{tikz}
\usetikzlibrary{shapes}
\usepackage{pgfplots}
	\pgfplotsset{compat=newest}
\usepackage{pgfplotstable}
\usepgfplotslibrary{fillbetween}
\usepackage{amsfonts}

\usepackage{bm}
\usepackage{mathtools}
\usepackage[utf8]{inputenc} 
\usepackage[T1]{fontenc}    
\usepackage{hyperref}       
\usepackage{url}            
\usepackage{booktabs}       
\usepackage{amsfonts}       
\usepackage{nicefrac}       
\usepackage{microtype}      
\usepackage{lipsum}		
\usepackage{graphicx}
\usepackage{doi}
\usepackage{makecell}

\theoremstyle{definition}

\theoremstyle{remark}

\newcommand{\fig}[4]{\begin{figure}[ht]\centering\includegraphics[width=#1\linewidth]{Figures/#2}\caption{#3}\label{#4}\end{figure}}

\newcommand\R[0]{\mathds{R}}
\newcommand\argmax[0]{\mathrm{argmax}}

\newcommand{\Cobs}[0]{\mathbf{C}_{\mathrm{obs}}}

\title{Uncertainty quantification in neutron and gamma time correlation measurements}

\author{\hspace{1mm}Paul Lartaud \\
        CEA DAM Île-de-France\\
        Centre de Mathématiques Appliquées, \\
        École polytechnique, \\
        Institut Polytechnique de Paris\\
	\texttt{paul.lartaud@polytechnique.edu} \\
	\And \hspace{1mm}Philippe Humbert \\
        CEA DAM Île-de-France\\
	\And \hspace{1mm}Josselin Garnier \\
        Centre de Mathématiques Appliquées, \\
        École polytechnique, \\
        Institut Polytechnique de Paris\\
}

\date{}

\renewcommand{\headeright}{}
\renewcommand{\undertitle}{}
\renewcommand{\shorttitle}{Uncertainty quantification in neutron and gamma time correlation measurements}

\hypersetup{
pdftitle={UQ-neutron-gamma-Lartaud},
pdfsubject={stat.AP},
pdfauthor={P. Lartaud, P. Humbert and J. Garnier},
pdfkeywords={sequential designs, surrogate model, inverse problem},
}

\begin{document}
\maketitle

\begin{abstract}
Neutron noise analysis is a predominant technique for fissile matter identification with passive methods. Quantifying the uncertainties associated with the estimated nuclear parameters is crucial for decision-making. A conservative uncertainty quantification procedure is possible by solving a Bayesian inverse problem with the help of statistical surrogate models but generally leads to large uncertainties due to the surrogate models' errors. 
In this work, we develop two methods for robust uncertainty quantification in neutron and gamma noise analysis based on the resolution of Bayesian inverse problems. We show that the uncertainties can be reduced by including information on gamma correlations. The investigation of a joint analysis of the neutron and gamma observations is also conducted with the help of active learning strategies to fine-tune surrogate models. We test our methods on a model of the SILENE reactor core, using simulated and real-world measurements.
\end{abstract}

\section{Introduction}
The identification of fissile matter is a foundational aspect of nuclear security and safeguards. It can be carried out through various experimental methods, each offering complementary insights. Among these, passive non-destructive assay techniques such as neutron time-correlation measurements and gamma spectroscopy are commonly employed \cite{BEDDINGFIELD1998405, dewji2016validation, bruggeman1996neutron}. This work focuses on neutron correlation measurements as a way to provide estimates of key nuclear parameters essential for material identification. However, these estimates often carry uncertainties that require thorough quantification for informed decision-making. The uncertainty quantification necessitates the resolution of an inverse problem, within a Bayesian framework. Sampling the posterior distribution of the nuclear parameters in a Bayesian inverse problem requires many calls to a computer model such as a Monte Carlo code for neutron transport. Because of the cost of these computer models, and the bias of analytical models like the point model \cite{pazsit2007neutron}, the resolution of the inverse problem relies on surrogate models, such as Gaussian Process (GP) models. 

In this paper, we explore two approaches for incorporating gamma time correlations into the inverse problem framework for fissile material identification. We demonstrate that integrating gamma measurements can significantly reduce uncertainties in the estimation of nuclear parameters. These methods are applied to both experimental and simulated data from the SILENE reactor. Furthermore, we investigate sequential design strategies as a means to optimize surrogate models, tailoring them to specific observational data within the inverse problem.

\section{Neutron correlations for fissile matter identification}
\subsection{Neutron time-correlations in the point model framework}
Throughout this paper, we focus on neutron noise analysis in zero-power systems. Neutron noise also refers to the study of neutron fluctuations in power reactors (see for example \cite{fry1971experience, demaziere2009numerical, ortiz2006bwr, ando1975void}), however, this is not the focus of this work.

The goal of neutron noise analysis is to identify some characteristics of the medium based on the fluctuations of the neutron population. It has been applied to supercritical systems during the start-up phase, at which point the neutron population is still low enough for the stochasticity to play a decisive role \cite{harris1965neutron}. Other works focused on the probability of extinction of the branching process describing the neutron population for such systems \cite{williams1979exact, tantillo2024new, williams2015time}. In \cite{cooling2016uncertainty}, the focus is on the uncertainty quantification of the estimated probability of extinction. For pulsed experiments without external sources, the intrinsic source term (either from $^{238}$U or $^{240}$Pu) is low and the burst time of the pulse can vary between two successive experiments because of the predominant stochastic fluctuations at the start-up. Numerous studies were conducted to estimate the behavior of the random burst times \cite{williams2016burst, authier2014initiation, humbert2004simulation, hansen1960assembly}.  

In this paper, our attention is directed at zero-power subcritical systems. The neutron fluctuations in zero-power systems depend on the multiplicity of the fission reactions and their analysis provides insights on the medium itself. For instance, it is common practice to infer the prompt decay constant $\alpha$ with the so-called Rossi-$\alpha$ method, which studies the probability of a second detection at $T+t$, given that a neutron was detected at $t$ \cite{rossi1941cosmic, hua2021fast, kuramoto2005rossi}. It is also possible to study the auto and cross power spectral densities, which are obtained by Fourier transform of the corresponding correlation function of the detector signals \cite{cohn1960simplified, pakari2018current, munoz2001subcritical, rugama2004experimental}. One may also analyze the distribution of the time intervals between two successive neutron detections. In a multiplying medium, this distribution differs from a memoryless exponential distribution, which occurs for standard non-multiplying nuclear reactions \cite{babala1967interval}.

In this paper, the main point of focus is the Feynman-$\alpha$ method first introduced in \cite{feynman1956dispersion}. This approach focuses on the variance-over-mean ratio of the detection distribution. In a non-multiplying medium, this ratio is equal to one, because the detection and source processes are Poisson processes, and Poisson distributions have equal mean and variance. However, in a multiplying medium, an excess of variance is introduced by the fission multiplicity. We will describe it in detail in the next paragraphs. 

The Feynman-$\alpha$ method is based on the estimation of the Feynman moment of the neutron count distribution. For a given time window of size $T$, the Feynman moment of order $n$ is denoted $Y_n(T)$. It is defined with respect to the binomial cumulants $\Gamma_n(T)$ of the distribution. The binomial cumulants represent the average number of combinations of $n$ correlated detections in a time $T$. Correlated detections refer to the simultaneous detections of multiple neutrons belonging to the same fission chain. Such neutrons are correlated in time, as opposed to neutrons from different initial source events which are independent. 

Consider the probability $p_n(t)$ of having detected $n$ neutrons at time $t$ given zero initial neutrons and a source term in the subcritical system given by a compound Poisson source (like a spontaneous fission source for example). The generating function $g(z, t)$ for the probabilities $p_n(t)$ is defined for $|z| < 1$ by:
\begin{equation}
    g(z, t) = \sum\limits_{n=0}^{+ \infty} p_n(t) z^n.
\end{equation}
The binomial cumulants $\Gamma_n$ of this distribution are given by:
\begin{equation}
    \Gamma_n(T) = \frac{1}{n!} \left( \frac{\partial ^n  \log g(z, t)}{\partial z^n} \right)_{z=1}.
\end{equation}
Finally, the Feynman moment of order $n$ is obtained by the following relation:
\begin{equation}
    Y_n(T) = \frac{n! \ \Gamma_n(T)}{\Gamma_1(T)}.
\end{equation}
In its original paper, Feynman used only the second order moment $Y_2(T)$ which will be denoted $Y(T)$ for the rest of the paper. It can be understood as a measure of the excess variance of the detection statistics compared to a Poisson distribution. 

Under the point model approximation, the Feynman moment of order $2$ can be analytically evaluated \cite{bell1965stochastic, pazsit2007neutron}. Let $\rho = \frac{k_p - 1}{k_p} < 0$ be the prompt reactivity where $k_p$ is the prompt multiplication factor. The prompt decay constant is denoted $\alpha$. $\varepsilon_F$ is the Feynman efficiency defined as the number of detections over the total number of induced fissions in the material. Consider a compound Poisson source which is a mix of a $(\alpha, n)$ source and a spontaneous fission source. Let $S$ be the source intensity expressed in source events per second and let $x_s$ be the fraction of source neutrons born by spontaneous fissions. Finally, we introduce the multiplicity parameters for the spontaneous and induced fissions. Let $\overline{\nu}$ be the average number of neutrons produced per induced fission and let $D_2$ and $D_3$ be the second and third-order Diven factor defined by $D_2 = \frac{\overline{\nu(\nu-1)}}{\overline{\nu}^2}$ and $D_3 = \frac{\overline{\nu(\nu-1)(\nu - 2)}}{\overline{\nu}^3}$, where $\nu$ is the random variable describing the fission multiplicity, and the bar stands for the average. Similarly, the quantities $\overline{\nu}_s$, $D_{2, s}$ and $D_{3, s}$ are introduced for the spontaneous fissions. Then the second Feynman moment is given by:
\begin{equation}\label{eq:feyn2}
    Y_2(T) = \frac{\varepsilon_F D_2}{\rho^2} \left(1 - x_s \rho \frac{\overline{\nu}_s D_{2, s}}{\overline{\nu} D_2} \right) \left(1 - \frac{1 - e^{- \alpha T}}{\alpha T} \right).
\end{equation}
The Feynman-$\alpha$ method involves evaluating the Feynman moment and fitting it to the previous expression to extract information on the kinetics of the system such as the prompt decay constant $\alpha$ or the reactivity $\rho$. Moreover, the point model approximation allows to derive an analytical relation for the average count rate $R$ which is independent of $T$:
\begin{equation}\label{eq:R1}
    R = - \frac{1}{x_s + \overline{\nu}_s - x_s \overline{\nu}_s} \frac{\varepsilon_F \overline{\nu}_{s} S}{\rho \overline{\nu}}.
\end{equation}
Finally, in \cite{furuhashi1968third}, the third Feynman moment was introduced and derived analytically, extending the method. The third moment is given by:
\begin{equation}\label{eq:feyn3}
   \begin{aligned}
    Y_3(T) = \ &3 \left(\frac{\varepsilon_F D_2}{\rho^2}\right)^2 \left(1 - x_s \rho \frac{\overline{\nu}_s D_{2, s}}{\overline{\nu} D_2} \right) \left(1 + e^{- \alpha T} - 2 \frac{1 - e^{- \alpha T}}{\alpha T} \right) \\
    &- \frac{\varepsilon_F^2 D_3}{\rho^3} \left(1 - x_s \rho \frac{\overline{\nu}_s^3 D_{3, s}}{\overline{\nu}^3 D_3} \right) \left(1  - \frac{3 - 4e^{- \alpha T} + e^{- 2\alpha T}}{2\alpha T} \right).
   \end{aligned}
\end{equation}
This work uses the notations $Y(T)$ for the second Feynman moment and $X(T)$ for the third Feynman moment as is done in \cite{endo2006space}. 

Our objective is to identify a fissile matter by estimating nuclear parameters such as the multiplication factor. Because measurements of $Y(T)$ and $X(T)$ are often noisy, we also seek to evaluate the reliability of our estimates. For that purpose, we propose a method based on the three quantities $R$, $Y(T)$, and $X(T)$ and whose goal is to estimate some parameters of interest of the medium, while providing uncertainty quantification on those results. The application will be restricted to the asymptotic values of the Feynman moments since we are not directly interested in the prompt decay constant $\alpha$. The asymptotic Feynman moments are denoted $Y_{\infty}$ and $X_{\infty}$ respectively. More details on the practical estimation of the Feynman moments are given in \ref{app:feyn_estimation}.

\subsection{Gamma noise}
In this paper, we seek to add the gamma noise to the problem to provide additional information and thus reduce the estimation uncertainties. Gamma correlations are generally less studied in the literature, mainly because of the uncertainties associated with the gamma multiplicity data. The average number of gamma produced per fission, denoted by $\overline{\mu}$, is far more uncertain than its neutron counterpart. Similarly, the gamma Diven factors $D_2^{(\gamma)}$ and $D_3^{(\gamma)}$ are not well-known. Some efforts have been made in this direction \cite{pazsit2009combined, murray2014measurement}, but these nuclear data still convey a large uncertainty. 

The study of gamma noise can be performed in a very similar fashion. One can define the gamma Feynman moments $Y_n^{(\gamma)}(T)$. In this work, we focus on the quantities $(R^{(\gamma)}, Y_\infty^{(\gamma)}, X_\infty^{(\gamma)})$ which are the analog of the neutron count rate and Feynman moments. These quantities can be evaluated with either sequential or triggered binning, in the same manner as their neutron counterparts. The estimates tend to be less noisy because we generally have more gamma events recorded (mainly due to the higher fission multiplicity). Within the point model framework (see \cite{pazsit2007neutron, enqvist2010sample}), analytical relations can be derived between $x^{(\gamma)}  = (k_p, S, x_s, M_\gamma, \varepsilon_\gamma)$ and $y^{(\gamma)} = (R^{(\gamma)}, Y_\infty^{(\gamma)}, X_\infty^{(\gamma)})$, where $\varepsilon_\gamma$ is the gamma efficiency defined by the average number of gamma detection per induced fission, and $M_\gamma$ is the gamma multiplication defined as the average number of gammas created by one neutron.


\subsection{Uncertainty quantification in neutron noise analysis}
\subsubsection{Bayesian inverse problem}
We continue this introduction with an overview of Bayesian inverse problems. Denote by $x$ some unknown material characteristics belonging to an input space $\mathcal{X} \subset \R^p$. Consider $N \geq 1$ neutron correlation observations $\mathbf{y} = (y_k)_{1 \leq k \leq N}$ where $y_k = (R, Y_\infty, X_\infty)_k$ is a vector of size $d=3$ containing observations of the three quantities of interest. Given these observations, we want to estimate the corresponding material characteristics $x$ and their uncertainties. The standard approach is to use a Bayesian framework in which $x$ is modeled as a random variable. We assume that $x$ is given by a prior probability distribution with density $p(x)$. The prior may incorporate expert knowledge or information from other measurements. In this work, the prior is non-informative, it is uniform on a bounded domain. Our goal is then to estimate the posterior distribution $p(x | \mathbf{y})$ of $x$ given $\mathbf{y}$. Assuming the observations are given by $y_k = f(x) + \varepsilon_k$ with $\varepsilon_k$ being independent Gaussian random variables with zero-mean and covariance $\Cobs$, the posterior density is obtained with Bayes' theorem:
\begin{equation}
    p(x | \mathbf{y}) \propto p(x) \left( (2 \pi)^d |\Cobs| \right)^{-1/2} \exp  \left(- \frac{1}{2} \sum\limits_{k=1}^N \|y_k - f(x) \|_{\Cobs}^2 \right)
\end{equation}
where $\|a\|_{\mathbf{C}}^2 = a^T \mathbf{C}^{-1} a$ for any positive-definite matrix $ \mathbf{C}$ of size $d \times d$ and for any vector $a \in \R^d$, and $|\Cobs|$ is the determinant of $\Cobs$.

This posterior distribution can be sampled by Markov Chain Monte Carlo (MCMC) methods. Having access to the posterior distribution is key for uncertainty quantification since we can evaluate any quantity of interest with it, such as the mean, variance, or probability of exceeding a given threshold.

\subsubsection{Gaussian process surrogate model}
The Bayesian paradigm described in the previous section displays one major flaw. It requires the so-called direct model $f$, serving as the link between inputs $x$ and outputs $y$. In most applications, this direct model is either unknown or represented by a costly computer model. One could also use the point model equations, but the strong foundational assumptions of the model introduce a bias that limits the reliability of the uncertainty quantification \cite{lartaud2023multi}. Since MCMC sampling requires a large number of calls, we need an alternate model that is both faster than computer codes such as MCNP \cite{goorley2012initial}, and more reliable than the point model. In the field of uncertainty quantification, a common practice is thus to use surrogate models, usually built with supervised learning methods and serving as emulators for a complex computer code.

In this work, we focus on Gaussian process (GP) models. These models come with a native uncertainty quantification: at every input $x$, a GP model $z$ provides a Gaussian predictive distribution $z(x)$:
\begin{equation}
    z(x) \sim \mathcal{N}\left(m(x), \mathbf{C}(x) \right)
\end{equation}
where $m(x) \in \R^d$ is the predictive mean and $\mathbf{C}(x)$ is a $d \times d$ covariance matrix. This predictive distribution is obtained by conditioning the GP distribution by some training data $(\mathbf{x}, \mathbf{z})$ where $\mathbf{x} = (x_i)_{1 \leq i \leq n}$ are the $n$ training inputs with $x_i \in \R^p$ and $\mathbf{z} = (z_i)_{1 \leq i \leq n}$ are the training outputs, with $z_i \in \R^d$. The creation of the training dataset by numerical simulation with the computer code MCNP is described in section \ref{sec:dataset_creation}.

We are not providing a detailed introduction to GP theory in this paper. The main aspects to keep in mind are the following. A GP model is determined by the user-defined mean and covariance functions. The training step consists of selecting the values of some hyperparameters $\theta$ embedded in the mean and covariance functions, by maximizing the probability of the training data $p(\mathbf{z} | \mathbf{x}, \theta)$. The predictive distribution at every input point $x$ is Gaussian and the mean vector $m(x)$ and covariance matrix $\mathbf{C}(x)$ are given by simple matrix operations. The GP model is thus significantly faster than our complex computer code $f$. For more details on GP theory, we refer to \cite{rasmussen2006gaussian, gramacy2020surrogates}.

In this work, we want a conservative estimate of the uncertainties. Thus, we wish to include in the Bayesian inverse problem both the uncertainty of the GP model and the uncertainty linked to the noise in the observations $y_k$. This can be done by considering the following statistical model:
\begin{equation}
    y_k = z(x) + \varepsilon_k = m(x) + \eta(x) + \varepsilon_k
\end{equation}
where $\eta(x) \sim \mathcal{N}\left(0, \mathbf{C}(x) \right)$ is independent of $\varepsilon_k$ because the model error is linked to the training of the model and is independent of the observations.
The new posterior density associated with this statistical model can be written as:
\begin{align}\label{eq:global_likelihood}
    p(x | \mathbf{y}) &\propto p(x) L(\mathbf{y} | x) \nonumber \\
    &\propto p(x) \left|\mathbf{C}(x) + \frac{1}{N} \Cobs  \right|^{-1/2} \nonumber \\
    &\times \exp\left(-\frac{1}{2} (\overline{y} - m(x))^T \left(\mathbf{C}(x) + \frac{1}{N} \Cobs \right)^{-1}(\overline{y} - m(x)) \right)
\end{align}
with $\overline{y} = \frac{1}{N}\sum\limits_{k=1}^N y_k \in \R^d$.

It encompasses both sources of uncertainty, from the model bias and the observation noise, and can be sampled by MCMC methods.

The objective of this paper is thus to provide this posterior distribution for the inputs $x^{(n, \gamma)} = (k_p, \varepsilon_F, S, x_s, M_\gamma, \varepsilon_\gamma)$ given some observations of the quantities $y^{(n, \gamma)} = (R^{(n)}, Y_\infty^{(n)}, X_\infty^{(n)}, R^{(\gamma)}, Y_\infty^{(\gamma)}, X_\infty^{(\gamma)})$.

\section{Neutron and gamma inverse problem}
In this section, we discuss the possibility of including gamma noise measurements in our inverse problem resolution, with two different approaches. Before describing these methods, we provide a brief description of the applicative case and the inverse problem observations. 

\subsection{Problem description}
\subsubsection{The SILENE experimental reactor}
The SILENE facility was an experimental reactor designed for pulsed experiments and subcritical multiplicity measurements and operated between 1974 and 2014 \cite{humbert2018simulation}. The reactor was designed to study criticality accidents occurring with fissile solutions.

The core is a cylindrical tank of $36$ cm outer diameter filled with highly enriched ($93$ wt. \% in $^{235}$U) uranyl nitrate. A control rod is located at the center of the core to avoid the initial power excursion when the fissile solution is pumped into the core. The core is placed in a large room with thick concrete walls. The reactor had three main operation modes. It was mainly used for pulsed experiments in which the center control rod is rapidly removed from the core to create a power excursion up to $1000$ MW. It was also possible to slowly remove the rod with an additional source to mimic the free evolution of a criticality accident. Finally, the reactor could also be operated in steady-state mode, with slow adjustments of the control rod. 

The internal control rod is either a boron rod with a reactivity worth of $5.8 \$$ or a cadmium rod with a reactivity worth of $4.1 \$$. The rod is inserted in a canal of $7$ cm of internal diameter at the center of the core. An external neutron source is often added below the core. The source is a $100$ mCi Am-Be source. The objective of the external neutron source is to limit the variance of the burst time due to stochastic fluctuations at low neutron populations. 

A schematic view of the SILENE core is presented in Figure \ref{fig:silene_MCNP_view}. The dotted region below the core represents the location of the Am-Be source.
\fig{0.8}{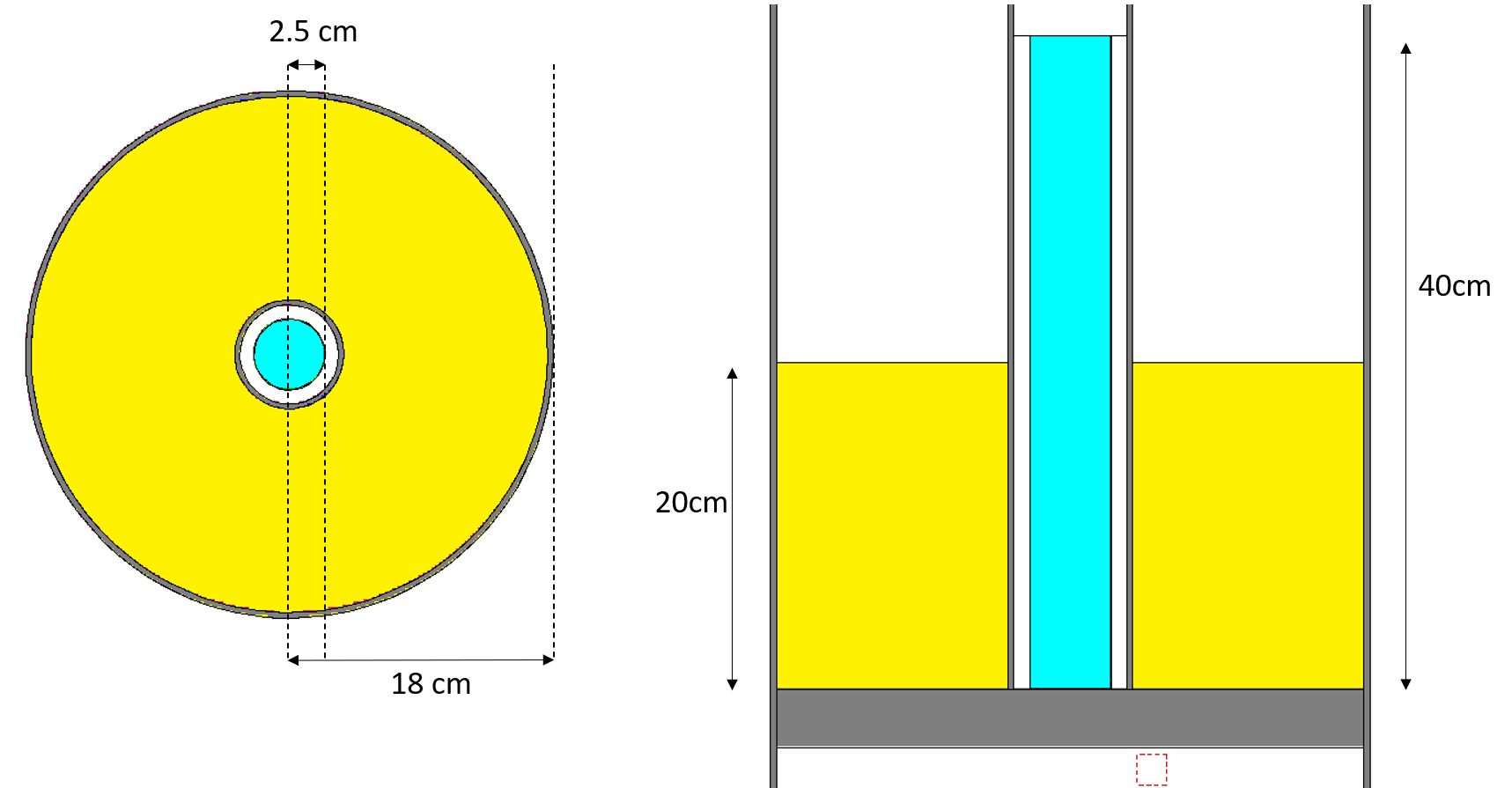}{Upper view (left) and side view (right) of the SILENE core as modeled in MCNP. The fissile solution is displayed in yellow, the steel in grey, and the detector in cyan.}{fig:silene_MCNP_view}

\subsubsection{Neutron and gamma observations}
We have access to a set of $N_n = 80$ time list files containing the detection instants of neutrons in the center detector, which was obtained during one of the measurement campaigns. These files can be post-processed with sequential binning to provide (independent) neutron observations $y_k^{(n)} = (R^{(n)}, Y_\infty^{(n)}, X_\infty^{(n)})_k$. 

However, we do not have access to gamma time list files. The gamma observations $(R^{(\gamma)}, Y_\infty^{(\gamma)}, X_\infty^{(\gamma)})$
are thus generated by MCNP simulations. In total, we produce $N_\gamma = 16$ gamma observations. In the MCNP simulations, we do not model a gamma detector. We simply record all gamma captures in the fissile region. This is a drastic simplification leading to large gamma efficiencies. The gamma observations are thus much less noisy than the neutron observations. However, the goal of this work is to demonstrate the feasibility of the joint neutron and gamma analysis. We do not seek to provide the best possible gamma model in this work, and this will be the subject of further studies and benchmarks.

\subsection{Sequential approach}
Our first method uses the Bayesian framework to leverage the additional knowledge brought by gamma correlations. Indeed, the prior distribution provides an easy way to incorporate new data in a Bayesian model. 

Consider some neutron observations $\mathbf{y}^{(n)} = (y_k^{(n)})_{1 \leq k \leq N_n}$ and some gamma observations $\mathbf{y}^{(\gamma)} = (y_k^{(\gamma)})_{1 \leq k \leq N_\gamma}$ for $N_n, N_\gamma \geq 1$. Throughout this paper, we recall the quantities of interest are $(R^{(n)}, Y_\infty^{(n)}, X_\infty^{(n)})$ for the neutron noise and $(R^{(\gamma)}, Y_\infty^{(\gamma)}, X_\infty^{(\gamma)})$ for the gamma noise. We assume the observations are related to some input $x^{(n)}$ and $x^{(\gamma)}$. Inspired by the point models for the neutron \cite{pazsit2007neutron} and gamma correlations \cite{pazsit2005calculation}, we define $x^{(n)} = (k_p, \varepsilon_F, S, x_s)$ and $x^{(\gamma)} = (k_p, S, x_s, M_\gamma, \varepsilon_\gamma)$. $x^{(n)}$ and $x^{(\gamma)}$ are not in the same design space. We introduce the neutron and gamma design spaces $\mathcal{X}^{(n)} \subset \R^4$ and $\mathcal{X}^{(\gamma)} \subset \R^5$.

We assume that GP surrogates are available for both the neutron and the gamma models. They are denoted respectively by $z^{(n)}$ and $z^{(\gamma)}$. They replace the unknown direct models such that the observations are:
\begin{equation}
    y_k^{(n)} = z^{(n)}(x^{(n)}) + \varepsilon^{(n)}_k \text{ for } 1 \leq k \leq N_n
\end{equation}
\begin{equation}
    y_k^{(\gamma)} = z^{(\gamma)}(x^{(\gamma)}) + \varepsilon^{(\gamma)}_k \text{ for } 1 \leq k \leq N_\gamma
\end{equation}
where $\varepsilon^{(n)}_k \sim \mathcal{N}\left(\mathbf{0}, \Cobs^{(n)} \right)$ and $\varepsilon^{(\gamma)}_k \sim \mathcal{N}\left(\mathbf{0}, \Cobs^{(\gamma)} \right)$ are the independent and identically distributed random variables describing the observational noise. For $x \in \mathcal{X}^{(n)}$ and $x^{(\gamma)} \in \mathcal{X}^{(\gamma)}$, we also introduce a neutron likelihood $L^{(n)}\left(\mathbf{y}^{(n)} | x^{(n)} \right)$ and a gamma likelihood $L^{(\gamma)}\left(\mathbf{y}^{(\gamma)} | x^{(\gamma)} \right)$ which are defined with the general likelihood \eqref{eq:global_likelihood}. In the sequential setting, we are thus dealing with two surrogate models in $d=3$ output dimension. 

To solve our inverse problem, we proceed as follows. Starting from a prior distribution with density $p(x^{(n)})$ for $x^{(n)} \in \mathcal{X}^{(n)}$, we solve the Bayesian inverse problem for the neutron correlations only, using the likelihood $L^{(n)}\left(\mathbf{y}^{(n)} | x^{(n)} \right)$. We thus obtain a posterior distribution with density $p^{(n)}\left(x^{(n)} | \mathbf{y}^{(n)}\right)$ for $x^{(n)} \in \mathcal{X}^{(n)}$. Since the posterior is obtained by MCMC sampling, its density is derived from the Markov chain samples by a Gaussian kernel density estimation. \\
Then, this posterior distribution is used as a prior in a second inverse problem where we introduce the gamma observations. However, because the neutron and gamma models have different input spaces the priors must be adjusted accordingly. One can notice that $x^{(n)}$ and $x^{(\gamma)}$ share three parameters: $k_p$, $S$ and $x_s$. Thus we define a prior on the parameters $(k_p, S, x_s)$ with the marginals of the posterior distribution $p^{(n)}\left(\ \cdot \ | \mathbf{y}^{(n)}\right)$. Then we affect a uniform prior for the two missing inputs $(M_\gamma, \varepsilon_\gamma)$. For $x^{(\gamma)} \in \mathcal{X}^{(\gamma)}$, the prior density for the gamma inverse problem, which is denoted by $p^{(\gamma)}(x^{(\gamma)})$ is then given by:
\begin{equation}
    p^{(\gamma)}(x^{(\gamma)}) = p(M_\gamma, \varepsilon_\gamma) \int_{\varepsilon_F} p^{(n)}(k_p, \varepsilon_F, S, x_s | \mathbf{y}^{(n)}) d\varepsilon_F
\end{equation}
where $p(M_\gamma, \varepsilon_\gamma)$ denotes the uniform prior density on a given subset of $\R^2$ for the inputs $M_\gamma$ and $\varepsilon_\gamma$. 

With this new prior distribution for the gamma inverse problem, we can compute a second posterior distribution $p^{(\mathrm{seq})}\left( \ \cdot \ | \mathbf{y}^{(\gamma)}, \mathbf{y}^{(n)} \right)$ whose density is given, for $x^{(\gamma)} \in \mathcal{X}^{(\gamma)}$ by:
\begin{equation}
    p^{(\mathrm{seq})}\left( x^{(\gamma)} | \mathbf{y}^{(\gamma)}, \mathbf{y}^{(n)} \right) \propto p^{(\gamma)}(x^{(\gamma)})  L^{(\gamma)}\left(\mathbf{y}^{(\gamma)} | x^{(\gamma)} \right).
\end{equation}
We highlight that this methodology is applicable in the reversed order, in which the gamma inverse problem is first solved before the neutron observations are added. \\
This final posterior distribution thus regroups the knowledge of neutron and gamma correlations. It requires surrogate models for both the neutron and gamma direct models. However, it has one main drawback: it treats the gamma and neutron problem sequentially and thus does not account for any correlations between the phenomena. This is a simplification since we expect the gamma and neutron models to be strongly correlated. The observations $\mathbf{y}^{(n)}$ and $\mathbf{y}^{(\gamma)}$ should not be treated independently from one another. This is why we describe a second approach in the next section. 

\subsection{Joint resolution}
To account for correlations between neutron and gamma observations, one could try to solve the joint inverse problem encompassing both types of observations. 

We define the joint input vector $x^{(n, \gamma)} = (k_p, \varepsilon_F, S, x_s, M_\gamma, \varepsilon_\gamma)$ and its design space $\mathcal{X}^{(n, \gamma)} \subset \R^6$. Let $p(x^{(n, \gamma)})$ be the uniform prior density on $\mathcal{X}^{(n, \gamma)}$ and let $N_{n, \gamma}$ be the number of observations for the joint inverse problem and $y^{(n, \gamma)} = (R^{(n)}, Y_\infty^{(n)}, X_\infty^{(n)}, R^{(\gamma)}, Y_\infty^{(\gamma)}, X_\infty^{(\gamma)})$ the vector containing both neutron and gamma observations. 

Consider a GP surrogate model $z^{(n, \gamma)}$ serving as an emulator for the joint direct model. The output dimension is now $d=6$ instead of $d=3$ for the sequential approach. The observations are given by the following statistical model:
\begin{equation}
    y^{(n, \gamma)}_k = z^{(n, \gamma)}\left(x^{(n, \gamma)}\right) + \varepsilon^{(n, \gamma)}_k
\end{equation}
for $1 \leq k \leq N_{n, \gamma}$ and where $\varepsilon^{(n, \gamma)}_k \sim \mathcal{N}\left(\mathbf{0}, \Cobs^{(n, \gamma)} \right)$ where $\Cobs^{(n, \gamma)}$ is obtained by the empirical covariance estimator applied to $\mathbf{y}^{(n, \gamma)}$. 

With this surrogate model, one can then define a likelihood $L^{(n, \gamma)}\left( \mathbf{y}^{(n, \gamma)} | x^{(n, \gamma)}\right)$ for $x^{(n, \gamma)} \in \mathcal{X}^{(n, \gamma)}$ with \eqref{eq:global_likelihood}. Finally, we obtain a posterior distribution $p^{(n, \gamma)}(\ \cdot \ | \mathbf{y}^{(n, \gamma)})$ which accounts for the correlations between neutron and gamma observations and whose density is given for $x^{(n, \gamma)} \in \mathcal{X}^{(n, \gamma)}$ by:
\begin{equation}
    p^{(n, \gamma)}(x^{(n, \gamma)} | \mathbf{y}^{(n, \gamma)} ) \propto L^{(n, \gamma)}\left(\mathbf{y}^{(n, \gamma)} | x^{(n, \gamma)} \right) p(x^{(n, \gamma)} ).
\end{equation}
Though this posterior distribution is theoretically more accurate, it requires a surrogate model of higher dimension, able to emulate both neutron and gamma correlations. The higher dimension for both the input and output spaces may hinder the predictive capabilities of this model, which would, in turn, make this approach less reliable.

\subsection{Building the models}
In this section, we investigate both of the approaches developed in the previous paragraphs. Consequently, we require a neutron surrogate model $z^{(n)}$  (\textbf{NSM}), a gamma surrogate $z^{(\gamma)}$ (\textbf{GSM}) and a joint surrogate model $z^{(n, \gamma)}$ (\textbf{JSM}) able to predict both neutron and gamma correlations. We will describe briefly these models in this section. 

\subsubsection{Training dataset creation}\label{sec:dataset_creation}
The first task of creating efficient surrogate models is to build training datasets. The datasets are created by MCNP simulations for both neutron and gamma measurements. Starting from the reference geometry of SILENE, we randomly change the composition, enrichment, source parameters, and geometry of the problem to produce new data instances. For each instance, an analog MCNP simulation is run to produce time list files for gamma and neutron detections. From these time list files, one can extract the measurements $(R^{(n)}, Y_\infty^{(n)}, X_\infty^{(n)}, R^{(\gamma)}, Y_\infty^{(\gamma)}, X_\infty^{(\gamma)})$. Moreover, we record, the inputs of interest $x^{(n, \gamma)} = (k_p, \varepsilon_F, S, x_s, M_\gamma, \varepsilon_\gamma)$. 
In this input vector, $k_p$ is obtained by an eigenvalue calculation, $M_\gamma$, $\varepsilon_F$, and $\varepsilon_\gamma$ are obtained by tally measurements. $S$ and $x_s$ are found directly in the MCNP input file and are changed randomly at each iteration. 

This process is iterated to produce datasets for neutron and gamma correlations. The datasets contain 232 instances.

\subsubsection{Model performance}
The two approaches described earlier require three different surrogate models. The sequential approach necessitates a gamma-only model, and a neutron-only model, while the joint approach needs a joint model predicting gamma and neutron outputs simultaneously. 

The \textbf{NSM} and \textbf{GSM} models are trained using the Coregionalization model for multi-output GP described in \cite{bonilla2007multi}. The Coregionalization model provides a covariance structure for a multi-output GP which can account for correlations across outputs. A simpler method would be to consider a single scalar GP for each output channel, however, this simplification impacts the reliability of the uncertainty quantification \cite{lartaud2023multi}. 

To further improve the models, the point model equations serve as the GP prior mean function. The hyperparameters are selected by maximization of the log-marginal likelihood, with the help of the bounded BFGS algorithm \cite{broyden1970convergence, fletcher1970new, goldfarb1970family, shanno1970conditioning}. $80$ \% of the dataset is used for training and the rest is kept as a test set to estimate model performance.  

The joint GP model is also an LMC model with 4 latent GPs, each having anisotropic Matérn kernels. However, due to the higher dimension, both for the input and output spaces, exact GP inference is too computationally expensive in this case. Indeed, we recall that the exact inference requires the inversion of a matrix of size $nd$ where $d$ is the output dimension, and $n=190$ is the number of training data. This amounts to a complexity $\mathcal{O}((nd)^3)$ which makes the inference intractable. 

Hence there is a need for sparse approximations to accelerate both training and inference. The sparse variational Gaussian process (SVGP) approach, introduced in \cite{titsias2009variational}, is used in this work. Our numerical developments are based on the \textit{GPyTorch} package developed for Python \cite{gardner2018gpytorch}. 
To evaluate the models' performance, we focus on the Normalized Mean Absolute Error (NMAE) and Normalized Root Mean Square Error (NRMSE). For $1 \leq i \leq n^{(*)}$, the $i$-th input and output samples in the test set are denoted respectively by $x^{(*, i)}$ and $z^{(*, i)}$. For the output $1 \leq j \leq d$, the NMAE and NRMSE are defined by:
\begin{equation}
    \mathrm{NMAE}_j = \frac{1}{n^{(*)}} \sum\limits_{i=1}^{n^{(*)}} \frac{ |m(x^{(*, i)})_j - z^{(*, i)}_j|}{\overline{z^{(*)}}_j}
\end{equation}
\begin{equation}
    \mathrm{NRMSE}_j = \sqrt{\frac{1}{n^{(*)}} \sum\limits_{i=1}^{n^{(*)}} \frac{\left(m(x^{(*, i)})_j - z^{(*, i)}_j\right)^2}{\overline{z^{(*)}}_j^2}}
\end{equation}
where $\overline{z^{(*)}}_j = \frac{1}{n^{(*)}} \sum\limits_{i=1}^{n^{(*)}} z^{(*, i)}_j$. We also introduce the $Q^2$ coefficient defined for the $j$-th output by:
\begin{equation}
    Q^2_j = 1 - \frac{\sum\limits_{i=1}^{n^{(*)}} \left(m(x^{(*, i)})_j - z^{(*, i)}_j\right)^2 }{\sum\limits_{i=1}^{n^{(*)}}\left(z^{(*, i)}_j - \overline{z^{(*)}}_j\right)^2}.
\end{equation}
The $Q^2$ highlights the fraction of the data variance explained by the regressor, the closer to one the better. 

Since we want to guarantee the reliability of the uncertainty quantification, we evaluate the coverage probabilities for different levels of confidence $\alpha \in (0, 1)$. In a $d$-dimensional setting, the coverage probability of confidence level $\alpha$, denoted by $C_p(\alpha)$ is defined by:
\begin{equation}
    C_p(\alpha) = \frac{1}{n^{(*)}} \sum\limits_{i=1}^{n^{(*)}} \mathds{1}_{I_\alpha(x^{(*, i)})} \left(z^{(*, i)} \right)
\end{equation}
where $I_\alpha(x)$ is the credible region of confidence level $\alpha$ defined by:
\begin{equation}
    I_\alpha(x) = \left\{y' \in \R^d  \ | \  \| y' - m(x) \|^2_{\mathbf{C}(x)}\leq q_{\alpha} \right\}
\end{equation}
and $q_\alpha$ is the quantile of level $\alpha$ of the $\chi^2$ distribution with $d$ degrees of freedom.

The error metrics are shown in Table \ref{tab:error_metrics_joint_model} for all models. One can see that the \textbf{NSM} and \textbf{GSM} models perform well. The errors on the gamma tend to be smaller due to the higher number of events in gamma correlations. However, the joint model \textbf{JSM} exhibits inferior predictive capabilities than these models. This can be explained by the higher dimensionality of the task for the \textbf{JSM} model, as well as by the sparse approximation. 
\begin{table}[ht]
\caption{Error metrics for the neutron model (\textbf{NSM}), gamma model (\textbf{GSM}) and joint model (\textbf{JSM}).}
\vspace{0.2cm}
\label{tab:error_metrics_joint_model}
\begin{minipage}{0.49\textwidth}
    \centering
    \begin{tabular}{|c|c c c|}
    \hline
    \textbf{NSM} & $\mathrm{NMAE}$ & $\mathrm{NRMSE}$ & $Q^2$ \\
    \hline
    $R$        & $0.008$ & $0.011$ & $0.9997$ \\
    $Y_\infty$ & $0.027$ & $0.038$ & $0.9983$ \\
    $X_\infty$ & $0.051$ & $0.153$ & $0.9919$ \\
    \hline
    \textbf{GSM} & $\mathrm{NMAE}$ & $\mathrm{NRMSE}$ & $Q^2$ \\
    \hline
    $R^{(\gamma)}       $ & $0.004$ & $0.006$ & $0.9999$ \\
    $Y_\infty^{(\gamma)}$ & $0.022$ & $0.031$ & $0.9953$ \\
    $X_\infty^{(\gamma)}$ & $0.080$ & $0.169$ & $0.9773$ \\
    \hline
    \end{tabular}
\end{minipage}
\begin{minipage}{0.49\textwidth}
    \centering
    \begin{tabular}{|c|c c c|}
    \hline
    \textbf{JSM} & $\mathrm{NMAE}$ & $\mathrm{NRMSE}$ & $Q^2$ \\
    \hline
    $R$        & $0.012$ & $0.016$ & $0.9990$ \\
    $Y_\infty$ & $0.046$ & $0.092$ & $0.9854$ \\
    $X_\infty$ & $0.089$ & $0.158$ & $0.9937$ \\
    \hline
    \textbf{JSM} & $\mathrm{NMAE}$ & $\mathrm{NRMSE}$ & $Q^2$ \\
    \hline
    $R^{(\gamma)}       $ & $0.012$ & $0.018$ & $0.9992$ \\
    $Y_\infty^{(\gamma)}$ & $0.016$ & $0.023$ & $0.9992$ \\
    $X_\infty^{(\gamma)}$ & $0.025$ & $0.047$ & $0.9992$ \\
    \hline
    \end{tabular}
\end{minipage}
\end{table}

We also looked at the coverage probabilities for all the models. They are plotted in Figure \ref{fig:cov_proba_joint}. Since the test set has only $42$ instances, the coverage probabilities cannot be very accurate, but one can still notice that all models provide reasonable coverage probabilities. While the \textbf{JSM} model does show some tendency to overestimate uncertainties, it maintains satisfactory coverage probabilities, which is particularly noteworthy given its higher output dimension.
\fig{0.75}{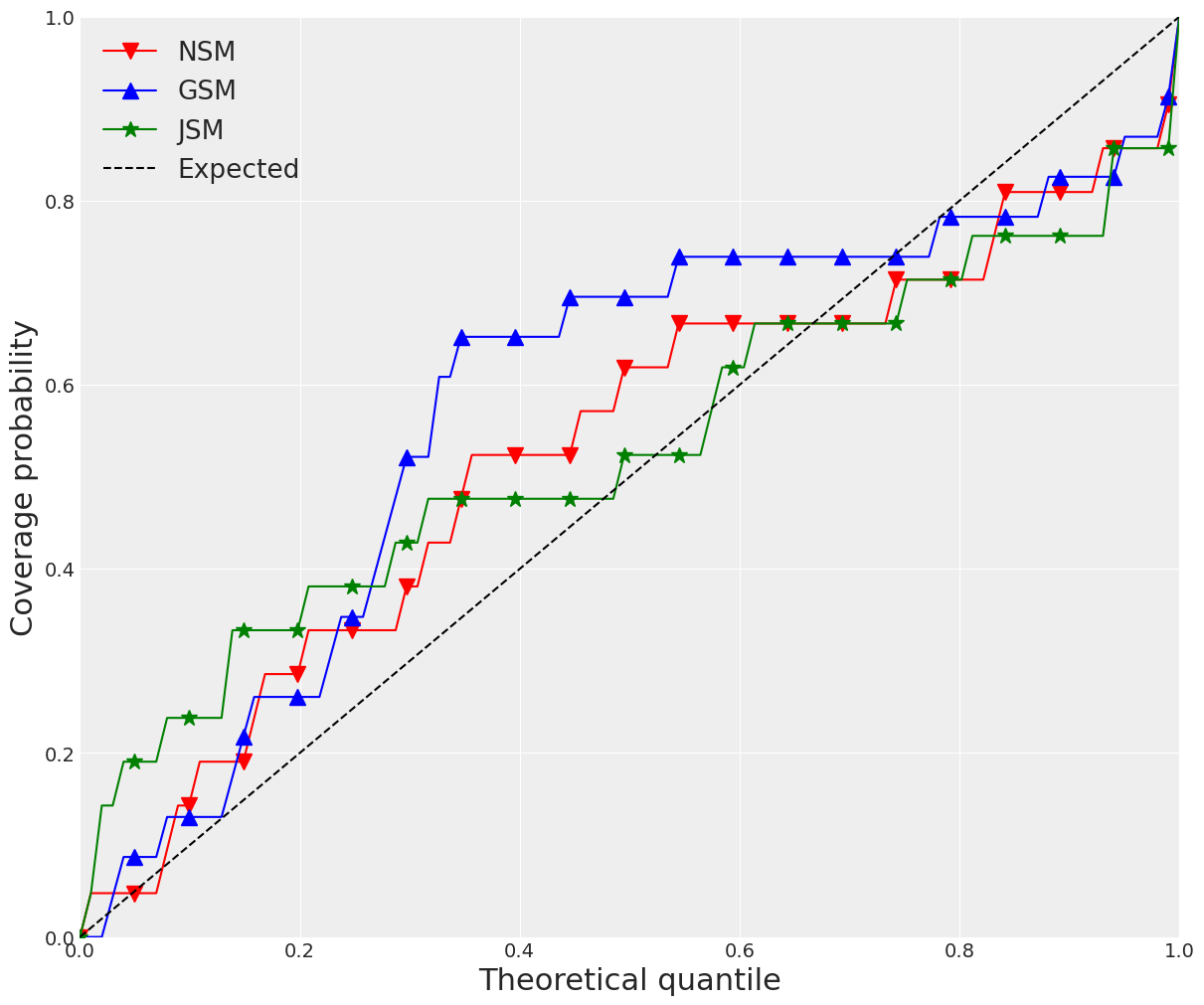}{Coverage probabilities evaluated on the test set for \textbf{NSM}, \textbf{GSM} and \textbf{JSM}.}{fig:cov_proba_joint}

\subsection{Application}
Our objective is now to apply the two proposed methods to experimental and simulated measurements from the SILENE reactor. We consider the configuration with a fissile height of $h = 20$ cm. We recall that the neutron observations $\mathbf{y}^{(n)}$ are obtained by post-treatment of experimental time list files. Since we do not have access to experimental gamma measurements, the gamma observations $\mathbf{y}^{(\gamma)}$ are obtained from numerical simulations with MCNP. We have a total of $N_\gamma = 16$ independent gamma observations.

The posterior distributions are obtained by MCMC sampling. The Adaptive Metropolis (AM) \cite{haario2001adaptive} algorithm is used for the sequential approach, in which the two inverse problems are solved one after the other. However, for the joint posterior distribution $p^{(n, \gamma)}(\ \cdot \ | \mathbf{y}^{(n, \gamma)} )$, we favor HMC-NUTS \cite{betancourt2017conceptual, hoffman2014no} due to the higher dimension of the problem, for which it is more suited. 

To highlight the gain in information brought by the gamma correlations, we plot in Figure \ref{fig:compare_gamma_joint_seq} the two-dimensional marginals for the inputs $(k_p, S)$ for the distributions $p^{(n)}\left(\ \cdot \ | \mathbf{y}^{(n)} \right)$, $p^{(\mathrm{seq})}(\ \cdot \ | \mathbf{y}^{(n, \gamma)} )$, $p^{(n, \gamma)}\left(\ \cdot \ | \mathbf{y}^{(n, \gamma)} \right)$.
\fig{0.99}{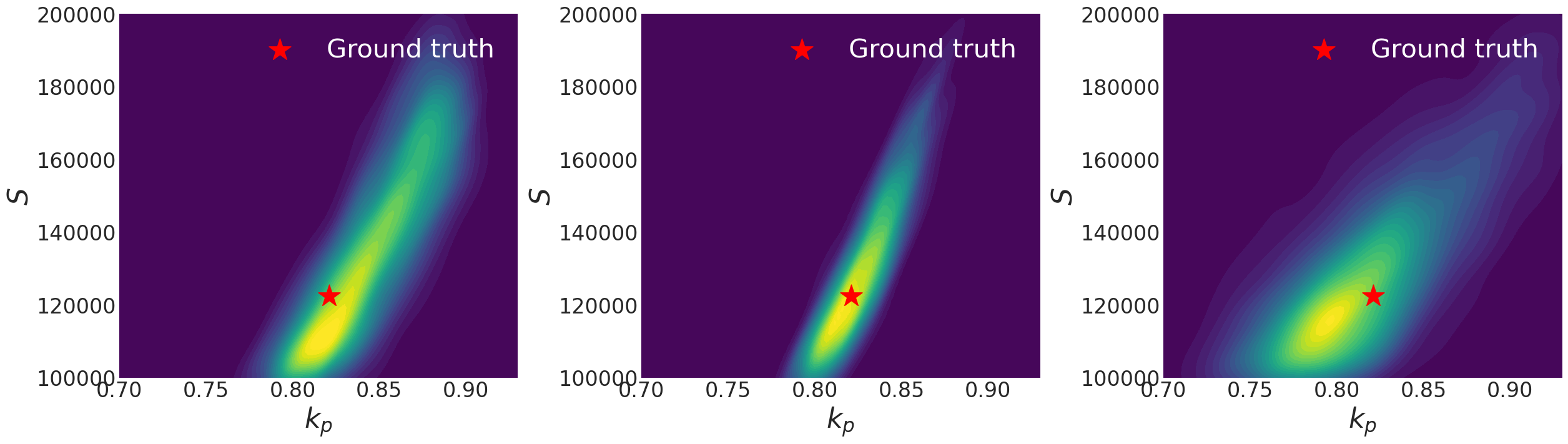}{Two-dimensional marginals w.r.t. $(k_p, S)$ of the posterior distributions obtained for the neutron inverse problem (left), the sequential approach (center) and the joint approach (right).}{fig:compare_gamma_joint_seq}
As one can see from this figure, the distribution is significantly narrower when we add the gamma correlations with the sequential approach. However, the joint approach yields a wider distribution. Looking back at the model performance in Table \ref{tab:error_metrics_joint_model} and Figure \ref{fig:cov_proba_joint}, the joint model does exhibit poorer prediction capabilities, which impacts the accuracy of the inverse problem resolution. However, we hope to resolve this limitation by efficiently adding more numerical data points to improve the GP surrogate model, with the help of a sequential design strategy.

\section{Sequential design strategy for the joint model}
\subsection{Sequential design for inverse problems}
Sequential design refers to strategies for the selection of new design points to improve a statistical model. Standard strategies try to leverage the knowledge of the model error to add training points in regions where the model uncertainty is the largest. For GP models, commonly used strategies are D-optimal designs which choose points maximizing the determinant of $\mathbf{C}(x)$ \cite{cook1980comparison}. Many other designs exist such as I-optimal designs or maximum entropy designs \cite{sacks1989designs}. We do not seek to provide an exhaustive list of all possible strategies. 

In our context, we wish to fine-tune the surrogate model \textbf{JSM} for the specific inverse problem studied. Ideally, the newly added training points should be selected close to the posterior distribution $p^{(n, \gamma)}(\ \cdot  \ | \mathbf{y}^{(n, \gamma)})$. For this reason, we adapt the D-optimal strategy to our problem by proposing a new strategy entitled Constraint Set Query (CSQ) \cite{lartaud2024sequential}. The CSQ strategy seeks to maximize the determinant of $\mathbf{C}(x)$ for $x$ in a subset $\mathcal{B}_h \subset \mathcal{X}^{(n, \gamma)}$ of the input space defined for $h \in \R^+$ by:
\begin{equation}
    \mathcal{B}_h = \left \{ x \in \mathcal{X}^{(n, \gamma)} | \log p^{(n, \gamma)}(x_m | \mathbf{y}) - \log p^{(n, \gamma)}(x | \mathbf{y}^{(n, \gamma)}) \leq h  \right\}
\end{equation}
where $x_m \in \argmax_{x \in \mathcal{X}^{(n, \gamma)}} \ p^{(n, \gamma)}(x | \mathbf{y}^{(n, \gamma)})$ is the maximum-a-posteriori (MAP). The newly selected training point $x^{(\mathrm{new})}$ is thus given by:
\begin{equation}\label{eq:opt_problem}
    x^{(\mathrm{new})} \in \argmax_{x \in \mathcal{B}_h} |\mathbf{C}(x)|.
\end{equation}
This strategy guarantees that $x^{(\mathrm{new})}$ is close to the MAP and avoids selecting points far away from the actual distribution. It is governed by a hyperparameter $h$ which is selected by the user. In this work, we have set $h=2$ allowing $x^{(\mathrm{new})}$ to spread quite far from the MAP. A lower $h$ constrains the point closer to the MAP. An intuitive look at the influence of $h$ can be obtained by considering a 1D Gaussian distribution $\mathcal{N}(\mu, \sigma^2)$. For this case, $\mathcal{B}_h = [\mu - \sqrt{2h} \sigma, \mu + \sqrt{2h} \sigma]$. For $h=2$, $\mathcal{B}_h$ is the $95 \%$ credible region. 

We apply the CSQ strategy iteratively to improve the \textbf{JSM} model with $20$ additional training points. The optimization problem \eqref{eq:opt_problem} is solved with the dual annealing algorithm implemented in \textit{scipy} \cite{xiang1997generalized}.

\subsection{Dealing with uncontrolled inputs}
One of the obstacles encountered while adapting the sequential design strategies to our case is the lack of control over the inputs $x$. Indeed, in a neutronic Monte Carlo simulation, $k_p$ is not a quantity that can be controlled. For simple geometries, one can expect how the $k_p$ might evolve when changing the problem, but we have no complete control over it. Similarly, $\varepsilon_F$ is an output of MCNP. How can we apply our strategies to actively improve the surrogate models in that context? 

We propose the following approach. Consider $x^{(\mathrm{new})} \in \mathcal{X}^{(n, \gamma)}$ the target design point recommended by the sequential design strategy. We introduce a cost function $L(x, x^{(\mathrm{new})})$ defined for $x \in \mathcal{X}^{(n, \gamma)}$ by:
\begin{equation}
    L(x, x^{(\mathrm{new})}) = \sum\limits_{j=1}^p \omega_j \left(x_j - x^{(\mathrm{new})}_j\right)^2
\end{equation}
where the $\omega_j \geq 0$ are weight factors such that $\sum\limits_{j=1}^p \omega_j = 1$ and where $p=6$ is the dimension of the input space. Ideally, we would like to reach $x=x^{(\mathrm{new})}$ which translates to $L(x, x^{(\mathrm{new})}) = 0$. Since we do not have complete control over $k_p$ and $\varepsilon_F$, reaching $x=x^{(\mathrm{new})}$ appears difficult, though we can try to minimize $L$. 

To do so, we proceed as follows. We start by running an MCNP simulation with $5 \times 10^4$ simulated neutrons, instead of the usual $5 \times 10^5$, to reduce the run time. This simulation yields an input vector $x$. Depending on the target $x^{(\mathrm{new})}$, we update the geometry and composition in the MCNP input file to produce a new $x$ which should be closer to $x^{(\mathrm{new})}$. For example, if $\varepsilon_F$ is lower than the target, we increase the radius of the detector region, or if $k_p$ is too large we reduce the enrichment. We iterate this process for $10$ iterations. The input file leading to the lower loss function $L(x, x^{(\mathrm{new})})$ is then used to create the new training instance. The weights $(\omega_j)_{1 \leq j \leq p}$ are selected using sensitivity analysis to estimate the respective impacts of the input parameters. The procedure is detailed in \ref{app:weight_selection}.

\subsection{Updated surrogate models}
Using the CSQ method, we update the joint model by adding $20$ new design points. Each new simulation is conducted with a total of $5 \times 10^5$ simulated neutrons. We present in Table \ref{tab:added_points} the mean relative error between the target design points and the design points obtained. 
\begin{table}[ht]
\vspace{0.2cm}
    \caption{Mean relative error between the target optimal design points given by CSQ and the actual points found with MCNP.}
    \label{tab:added_points}
    \renewcommand{\arraystretch}{1.}
    \centering
    \begin{tabular}{|c|c c c c|}
    \hline
     & $k_p$ & $\varepsilon_F$ & $M_\gamma$ & $\varepsilon_\gamma$ \\
    \hline
    Rel. error (\%) & $3.6$ & $71$ & $63$ & $40$\\
    \hline
    \end{tabular}
\end{table}

The lack of control in the selection of the inputs is significant, with some parameters such as $\varepsilon_F$ and $M_\gamma$ displaying a relative error superior to $50 \%$. However, we manage to get a satisfying precision on $k_p$ which is one of the main contributors to the joint inverse problem.

\begin{table}[ht]
\footnotesize
    \caption{Error metrics for the previous \textbf{JSM} and the updated \textbf{JSM}.}
\vspace{0.2cm}
    \label{tab:improved_jsm_metrics}
\begin{minipage}{0.45\textwidth}
    \centering
    \begin{tabular}{|c|c c c|}
    \hline
    \textbf{JSM} & $\mathrm{NMAE}$ & $\mathrm{NRMSE}$ & $Q^2$ \\
    \hline
    $R$        & $0.012$ & $0.016$ & $0.9990$ \\
    $Y_\infty$ & $0.046$ & $0.092$ & $0.9854$ \\
    $X_\infty$ & $0.089$ & $0.158$ & $0.9937$ \\
    \hline
    \textbf{JSM} & $\mathrm{NMAE}$ & $\mathrm{NRMSE}$ & $Q^2$ \\
    \hline
    $R^{(\gamma)}       $ & $0.012$ & $0.018$ & $0.9992$ \\
    $Y_\infty^{(\gamma)}$ & $0.016$ & $0.023$ & $0.9992$ \\
    $X_\infty^{(\gamma)}$ & $0.025$ & $0.047$ & $0.9992$ \\
    \hline
    \end{tabular}
\end{minipage}
\begin{minipage}{0.54\textwidth}
    \renewcommand{\arraystretch}{1.}
    \centering
    \begin{tabular}{|c|c c c|}
    \hline
    Updated \textbf{JSM} & $\mathrm{NMAE}$ & $\mathrm{NRMSE}$ & $Q^2$ \\
    \hline
    $R$        & $0.007$ & $0.009$ & $0.9997$ \\
    $Y_\infty$ & $0.035$ & $0.045$ & $0.9965$ \\
    $X_\infty$ & $0.096$ & $0.317$ & $0.9337$ \\
    \hline
    Updated \textbf{JSM} & $\mathrm{NMAE}$ & $\mathrm{NRMSE}$ & $Q^2$ \\
    \hline
    $R^{(\gamma)}       $ & $0.009$ & $0.013$ & $0.9996$ \\
    $Y_\infty^{(\gamma)}$ & $0.016$ & $0.023$ & $0.9992$ \\
    $X_\infty^{(\gamma)}$ & $0.030$ & $0.058$ & $0.9988$ \\
    \hline
    \end{tabular}
\end{minipage}
\end{table}
The new metrics for the improved joint model are shown in Table \ref{tab:improved_jsm_metrics}. The coverage probabilities are nearly unchanged.\\
To highlight the reduction in the epistemic uncertainty in the model, we also compared the determinants of the predictive covariances averaged over the test set (MCD). For a surrogate model with predictive distribution $z(x) \sim \mathcal{N}\left(m(x),\mathbf{C}(x) \right)$, this quantity is defined by:
\begin{equation}
    \mathrm{MCD} = \frac{1}{n^{(*)}} \sum\limits_{i=1}^{n^{(*)}} | \mathbf{C}(x^{(*, i)})|
\end{equation}
where we recall that $x^{(*, i)}$ is an instance of the test set of size $n^{(*)}$. 

The ratio of these quantities, obtained for the previous joint model and the updated model is $\frac{\mathrm{MCD}_{\mathrm{old}}}{\mathrm{MCD}_{\mathrm{new}}} \simeq 2.3$. On average, the determinants of the predictive covariances are reduced by a factor of more than $2$. Besides, thanks to the CSQ strategy, this uncertainty reduction is likely more prominent in regions of high posterior density.

\subsection{Improved posterior distribution}
With the newly enriched \textbf{JSM}, we can sample a new joint posterior distribution with HMC-NUTS. The marginal distribution for $(k_p, S)$ is displayed in Figure \ref{fig:gamma_neutron_updated} (right) along with the one obtained with the old \textbf{JSM} (left).
\fig{0.99}{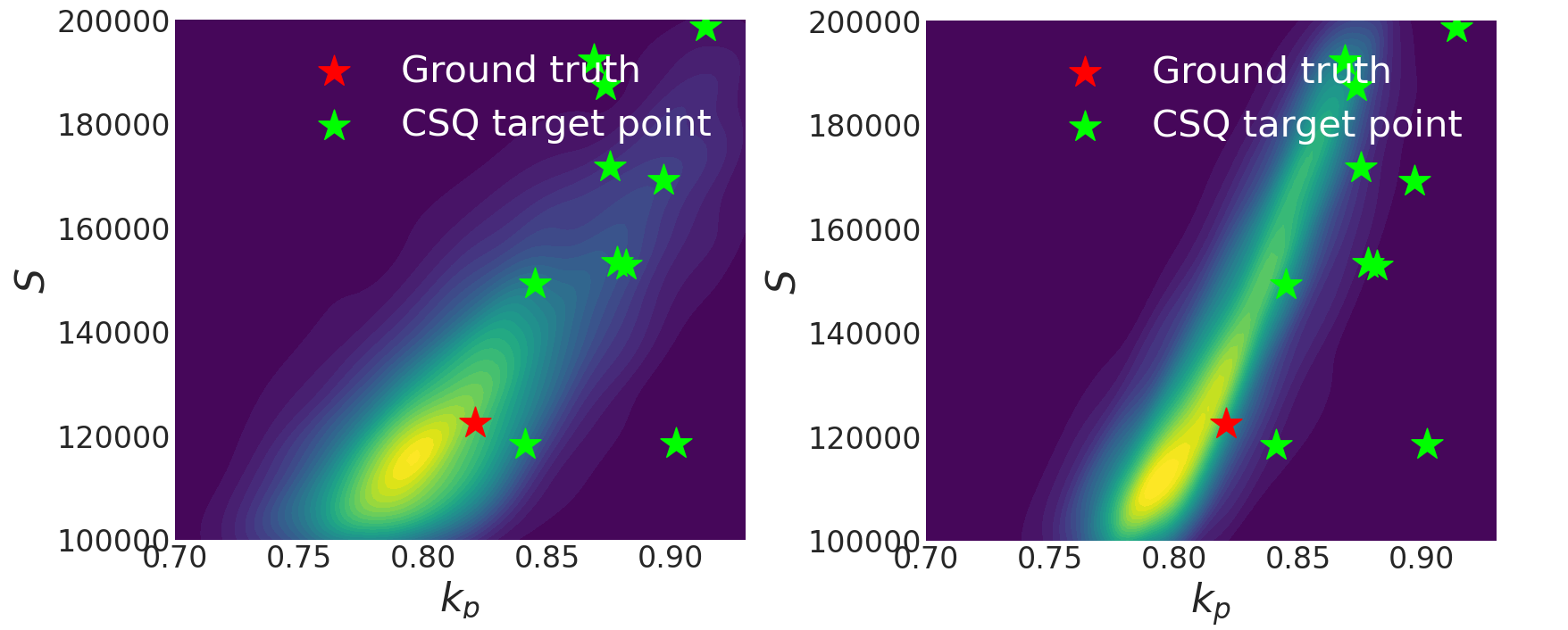}{Two-dimensional marginals w.r.t. $(k_p, S)$ obtained with the \textbf{JSM} surrogate (left) and the updated \textbf{JSM} surrogate (right). The first ten design points obtained by CSQ are also displayed.}{fig:gamma_neutron_updated}
The newly obtained posterior distribution shows a drastic improvement compared to the previous one. The posterior distribution remains more spread out than the one obtained with the sequential approach, however this may be caused by an underestimation of the uncertainties in the sequential approach. Indeed, the successive treatment of two inverse problems acts as if the neutron and gamma training data were uncorrelated. This is not true, as both the gamma and neutron Feynman moments increase with higher multiplication. Thus, there is some shared information between the neutron and gamma training data which is neglected in the sequential approach. On the other hand, the joint approach includes correlations between the neutron and gamma training data and is thus more reliable. 

With the help of sequential design strategies, the joint approach produces a reliable posterior distribution which reduce the uncertainties in the estimation of the nuclear parameters of interest. The designs are still hindered by the lack of control over the inputs in the design space and possibly by the sparse approximation in the \textbf{JSM} model. Developments in this direction would benefit the robustness of this approach.

\section{Conclusion}
\label{sec:conclusion}
In this paper, we explored gamma correlations and their potential to diminish uncertainties in the estimation of unknown nuclear parameters. We successfully demonstrated the potential gain in incorporating gamma correlations in the resolution of the inverse problem involved in the identification of fissile matter. The inclusion of gamma correlation measurements can be done with either a sequential approach, in which the neutron and gamma inverse problems are solved sequentially, or jointly within a common framework. The sequential treatment of neutron and gamma correlations is shown to significantly decrease the uncertainties in the estimation of key nuclear parameters. Although theoretically better, the joint treatment is, at first, hindered by a lack of numerical training data for the underlying higher-dimensional surrogate model. For a comprehensive joint inverse problem, the selection of numerical design points becomes crucial in fine-tuning surrogate models and can be carried out with sequential design strategies adapted to Bayesian inverse problems. Despite the lack of control over the input data, these designs largely reduce the spread of the posterior distribution in the joint approach. The uncertainty quantification is more reliable than that of the sequential approach and proves the feasibility of including gamma correlation measurements in the identification of fissile matter. 

The integration of gamma correlations in the identification of fissile matter is promising but requires further development to ensure the approach's robustness. Notably, this work did not incorporate uncertainties in nuclear data. The evaluation of neutron multiplicity data has seen significant improvements in recent years, for example with the developments of new fission models such as FREYA \cite{verbeke2015fission}. The gamma multiplicity on the other hand has received less attention, necessitating further development to achieve a comprehensive and reliable uncertainty quantification framework for scenarios involving gamma correlations.
Additionally, gamma correlations are highly sensitive to uncertainties in nuclear data due to the higher detection event rates. While gamma observations may appear to exhibit lower noise levels, hidden uncertainties in the nuclear data could be present. Moreover, this study was based exclusively on numerical simulations of gamma correlation measurements, highlighting the necessity of validating the proposed approach with real-world observations to ensure reliable uncertainty quantification.

\pagebreak       
\bibliographystyle{apalike}   
\bibliography{bibliography}
\newpage
\appendix
\section{Feynman moments estimation}\label{app:feyn_estimation}
The estimation of the Feynman moments of second and third order is crucial for our problem. Two methods are presented here, the sequential binning approach which is the most straightforward, and the filtered binning method which makes use of the additional information on the fission chains provided by Monte Carlo codes. 

\subsection{Sequential binning}
Let us consider first the sequential binning approach. Our objective is to estimate the Feynman moments for a given time window $T$ given the PTRAC file, which records all the detection times. Let $T_{\mathrm{max}}$ be the end time of the experiment. For a Monte Carlo simulation, this end time is artificial since we simulate several neutrons individually without a global time scale. However, the global time scale is brought by the chosen arbitrary source intensity $S$. In sequential binning, the whole experiment is split into $W = \lfloor \frac{T_{\mathrm{max}}}{T} \rfloor \in \mathds{N}$ time windows of size $T$ numbered from $1$ to $W$. Let $(n_w)_{1 \leq w \leq W}$ be the number of neutrons detected in each window $w$. The Feynman moment are linked to the ordinary moments $(M_p)_{p \geq 1}$ of this count statistics which can be estimated for $p \geq 1$ by the standard ordinary moment estimator $\widehat{M_p}$ defined as:
\begin{equation}
    \widehat{M_p} = \frac{1}{W} \sum\limits_{w=1}^W n_w^p.
\end{equation}
Now the Feynman moments can then be found by the following estimators.
\begin{equation}
    \widehat{Y(T)} = \frac{\widehat{M_2}}{\widehat{M_1}} -\widehat{M_1} - 1
\end{equation}
\begin{equation}
    \widehat{X(T)} = \frac{\widehat{M_3}}{\widehat{M_1}} + 2\left(\widehat{M_1}^{2} + 1\right) - 3 \left( \frac{\widehat{M_2}}{\widehat{M_1}} + \widehat{M_2} - \widehat{M_1} \right).
\end{equation}
Since we are interested in the asymptotic values of the Feynman moments we would like to plot their evolution with $T$ to guarantee the convergence towards the asymptotic value. Hence, the time windows are merged two by two and we then proceed similarly to obtain an estimation of $Y(2T)$ and $X(2T)$. Iteratively, we can obtain estimations for the Feynman moments for all $nT$ with $n \geq 1$. Of course, the larger $n$ the fewer time windows, leading to more noisy estimations. A compromise has to be made in that regard. 

\subsection{Filtered triggered binning}
In filtered triggered binning, we leverage the additional information provided by the numerical simulation. In this paper, we are working with the neutronic Monte Carlo code MCNP6. Since neutrons are simulated one by one, for each detection we know the corresponding source neutron. Each detection belongs to a fission chain that can be identified. Since the Feynman moments are the average relative number of double and triple-correlated detections, this additional information allows us to remove accidental correlations occurring due to random chance. Thus, the noise in the estimations with sequential binning can be greatly removed. 

To use the information on the history number of the detection, we proceed as follows. At each neutron detection occurring at an instant $t_k$, a time window with size $T$ is opened. This initial detection event is called the triggering event. In this window, the detections between $[t_k, t_k + T]$ are recorded if and only if they are truly correlated to the triggering event, or in other words if they belong to the same fission chain. This can be easily checked as MCNP6 provides the history number for each of the recorded events. 

Let $n^{(t)}_k$ be the number of detected neutrons in the triggered window. Then the average number of double and triple correlated detections denoted respectively $N_{2c}(T)$ and $N_{3c}(T)$ can be estimated by:
\begin{equation}
    N_{2c}(T) \simeq \widehat{N_{2c}(T)} = \frac{1}{N_{\mathrm{det}}} \sum\limits_{k=1}^{N_{\mathrm{det}}} n^{(t)}_k
\end{equation}
\begin{equation}
    N_{3c}(T) \simeq \widehat{N_{3c}(T)} = \frac{1}{N_{\mathrm{det}}} \sum\limits_{k=1}^{N_{\mathrm{det}}} \frac{n^{(t)}_k (n^{(t)}_k - 1)}{2}
\end{equation}
where $N_{\mathrm{det}}$ is the total number of detections. From these, the Feynman moments can be estimated using:
\begin{equation}
    Y(T) \simeq \widehat{Y(T)} = \frac{2 \widehat{N_{2c}(T)}}{N_{\mathrm{det}}}
\end{equation}
\begin{equation}
    X(T) \simeq \widehat{X(T)} = \frac{6 \widehat{N_{3c}(T)}}{N_{\mathrm{det}}}
\end{equation}
These triggered binning estimators allow us to estimate the Feynman moments while filtering out the noise due to the accidental correlations. However, the latter can only be used in numerical simulations, as for practical experiments the history number of a given detection signal cannot be recovered. Thus, for this work, the triggered binning is only used to generate the data that will be used to train our models. To solve the inverse problems, the observations considered are based on sequential binning estimations, such that they mimic real-world experiments.

\section{Selection of the weights \texorpdfstring{$\omega_j$}{wj}}\label{app:weight_selection}
The weights $(\omega_j)_{1 \leq j \leq p}$ should be defined to account for the influence of the variables. Our first attempt was to define arbitrary weights based on our knowledge of the problem. While this may work, we refined our method by using sensitivity analysis to better select the weights. The sensitivity analysis study is performed using the Analysis of Covariance (ANCOVA) approach.

Consider a scalar function $f$ of a random variable $x = (x_1, ..., x_p) \in \mathcal{X}$ for $p \geq 1$. The model response can be decomposed into a sum of contributions such as:
\begin{equation}\label{eq:hoeffding}
    f(x) = f_0 + \sum\limits_{i=1}^p f_i(x_i) + \sum_{1 \leq i, j \leq p} f_{i, j}(x_i, x_j) + ... = \sum\limits_{A \in \mathcal{P}_p} f_A(x_A)
\end{equation}
where $\mathcal{P}_p$ is the set of subsets of $\left\{1, ..., p \right\}$, $x_A$ is the subvector of $x$ where only the components in $A$ are kept, $f_A(x_A) = \sum\limits_{B \in \mathcal{P}_A} (-1)^{|A| - |B|} \mathbb{E}\left[ f(x) | x_B \right]$, and $f_0 = \mathbb{E} \left[ f(x) \right]$. In this second expression, $\mathcal{P}_A$ is the set of subsets of $A$, and $| \cdot |$ refers to the cardinality of the set.

This result is known as the Hoeffding decomposition \cite{sudret2008global, da2021basics}. In the case of independent inputs, the terms of the decomposition are orthogonal. The sensitivity analysis of the model response $f(x)$ can then be performed using an ANOVA (analysis of variance) decomposition of $f(x)$ such that:
\begin{equation}
    \mathrm{Var}[f(x)] = \sum_{i=1}^p V_i + \sum\limits_{1 \leq i, j \leq p} V_{i, j} + ... = \sum\limits_{A \in \mathcal{P}_p} V_A
\end{equation}
where $V_A = \mathrm{Var}\left[f_A(x_A) \right]$.

In the case of dependent inputs, we lose the orthogonality of the terms in the Hoeffding decomposition. The variance is then written as:
\begin{equation}
    \mathrm{Var}[f(x)] = \sum\limits_{A \in \mathcal{P}_p} \mathrm{Cov}\left[f_A(x_A), f(x) \right]
\end{equation} 
\begin{table}[ht]
\caption{First-order structural Sobol indices.}
\vspace{0.2cm}
\label{tab:new_sobol_gamma}
\footnotesize
    \centering
    \begin{tabular}{|c|c c c | c c c | }
    \hline
    $s_{j, i}$ & $R^{(n)}$ & $Y_\infty^{(n)}$ & $X_\infty^{(n)}$ & $R^{(\gamma)}$ & $Y_\infty^{(\gamma)}$ & $X_\infty^{(\gamma)}$ \\
    \hline
    $k_p$                 & $4.3\times 10^{-1}$ & $5.7\times 10^{-1}$ & $6.6\times 10^{-1}$ & $4.1\times 10^{-4}$  & $2.6\times 10^{-1}$ & $2.9\times 10^{-1}$\\
    $\varepsilon_F$       & $6.8\times 10^{-1}$ & $3.9\times 10^{-1}$ & $7.4\times 10^{-1}$ & $1.9\times 10^{-6}$  & $2.1\times 10^{-5}$ & $2.7\times 10^{-6}$\\
    $S$                   & $5.7\times 10^{-2}$ & $1.1\times 10^{-4}$ & $7.2\times 10^{-4}$ & $4.9\times 10^{-2}$  & $4.4\times 10^{-5}$ & $2.2\times 10^{-5}$\\
    $x_s$                 & $6.5\times 10^{-2}$ & $3.2\times 10^{-3}$ & $1.5\times 10^{-3}$ & $5.0\times 10^{-2}$  & $2.5\times 10^{-3}$ & $8.7\times 10^{-4}$\\
    $M_\gamma$            & $5.3\times 10^{-2}$ & $2.6\times 10^{-2}$ & $1.9\times 10^{-1}$ & $1.8\times 10^{-1}$  & $1.2\times 10^{-1}$ & $3.5\times 10^{-1}$\\
    $\varepsilon_\gamma$  & $7.2\times 10^{-2}$ & $7.9\times 10^{-3}$ & $2.2\times 10^{-2}$ & $4.8\times 10^{-1}$  & $5.3\times 10^{-1}$ & $4.7\times 10^{-1}$\\
    \hline
    \end{tabular}
\end{table}

The structural Sobol index \cite{hart2018approximation} associated to the subset $A \in \mathcal{P}_p$ is defined by:
\begin{equation}
    s_A = \frac{\mathrm{Var}\left[f_A(x_A) \right]}{\mathrm{Var}\left[f(x) \right]}.
\end{equation}
In particular, the first-order Sobol index associated with the $j$-th input is $s_{\{j\}}$. It represents the contribution of the $j$-th input to the model output, not accounting for the correlation between the $j$-th input and the other inputs. These Sobol indices allow us to identify quantitatively the main contributors. For our application, we denote by $s_{j, i}$ the uncorrelated (or structural) first-order Sobol index associated with the $j$-th input of $x^{(n, \gamma)}$ and the $i$-th output of $y^{(n, \gamma)}$. They are estimated using a Polynomial Chaos Expansion \cite{sudret2013analysis} and are presented in Table \ref{tab:new_sobol_gamma}.

Intuitively, one could increase the contributions of the outputs whose observational variances are low, since the inverse problem is more sensitive to these outputs. We choose the non-normalized weights $\omega_j$ defined by:
\begin{equation}\label{eq:weights_uncontrolled}
    \omega_j = \sum\limits_{i=1}^d s_{j, i} \frac{\overline{y^2}_i }{\sigma_i^2}
\end{equation}
where $\sigma_i^2$ is the observational variance of the $i$-th output, for $1 \leq i \leq d$, and $\overline{y^2}_i = \frac{1}{N_{n, \gamma}} \sum\limits_{k=1}^{N_{n, \gamma}} \left( y^{(n, \gamma)}_k\right)_i^2$. We then normalize the weights $\omega_j$ to have $\sum_{j=1}^p \omega_j = 1$. 

With this empirical approach, we hope to put more weight on the outputs $R^{(n)}$ and $R^{(\gamma)}$ which exhibit low relative variance.   
\newpage
\end{document}